\documentclass[aps,prd,twocolumn,floatfix,superscriptaddress,showpacs,nofootinbib]{revtex4}
\usepackage[latin1]{inputenc}  
\usepackage{bbm,slashed}
\usepackage{mathrsfs}
\usepackage{graphicx,epsfig}
\graphicspath{{figures/}}
\usepackage{amsmath,amsfonts,amssymb} 
\usepackage{booktabs}

\usepackage{epstopdf}   
   
\usepackage{url}
 
\usepackage{dsfont}  
\usepackage{bm,bbm,xcolor} 
        
\usepackage{color}

\newcommand{\eq}[1]{(\ref{#1})}

\def\s0#1#2{\mbox{\small{$ \frac{#1}{#2} $}}}

\newcommand{\uas}{u_*}  
\newcommand{\psib}{\bar\psi}      
\newcommand{\ha}{\frac{1}{2}} 
\newcommand{\fdi}{\slashed{\partial}}
 
\newcommand{\cD}{\mathcal{D}}

\newcommand{\abs}[1]{\left| #1 \right|}

\DeclareMathOperator{\Str}{Str}

 \DeclareMathAlphabet{\boldmathe}{T1}{cmr}{bx}{it}

\begin{document}

\title{Critical behavior of supersymmetric $O(N)$ models in the large-$N$ limit}

\author{Daniel F. Litim}
\affiliation{Department of Physics and Astronomy, University of Sussex, BN1 9QH, Brighton, UK.}
\author{Marianne C. Mastaler}
\affiliation{ Theoretisch-Physikalisches Institut, Friedrich-Schiller-Universit{\"a}t
Jena,
Max-Wien-Platz 1, D-07743 Jena, Germany}
\author{Franziska Synatschke-Czerwonka}
\affiliation{ Theoretisch-Physikalisches Institut, Friedrich-Schiller-Universit{\"a}t
Jena,
Max-Wien-Platz 1, D-07743 Jena, Germany}
\author{Andreas Wipf}
\affiliation{ Theoretisch-Physikalisches Institut, Friedrich-Schiller-Universit{\"a}t
Jena,
Max-Wien-Platz 1, D-07743 Jena, Germany}

\begin{abstract}{
We derive a supersymmetric renormalization group (RG) equation 
for the scale-dependent superpotential  of the  supersymmetric $O(N)$
model in three dimensions. For a supersymmetric optimized regulator function
we solve the RG equation for the superpotential exactly in the large-$N$ limit. 
The fixed-point solutions are classified by an exactly marginal coupling. 
In the weakly coupled regime there exists a unique fixed point solution,
for intermediate couplings we find two separate fixed point solutions
and in the strong coupling regime no globally defined fixed-point potentials 
exist. We determine the exact critical exponents both for the superpotential 
and the associated scalar potential. Finally we relate the 
high-temperature limit of the four-dimensional theory to 
the Wilson-Fisher fixed point of the purely scalar theory.}
\end{abstract}

\maketitle

\section{Introduction} 

Fixed points of the renormalization group (RG) play a fundamental role in
statistical physics and quantum field theory \cite{Wilson:1971bg,Wilson:1971dh}.
Infrared (IR) fixed points dominate the long-distance behavior of correlation
functions and are relevant for the understanding  of continuous phase
transitions and universal scaling laws \cite{ZinnJustin}. Ultraviolet (UV)
fixed points control the short-distance behavior of quantum field theories. It
is widely believed that the existence of an UV fixed point
is mandatory for a definition of quantum field theory on a microscopic level, e.\,g.~asymptotic
freedom of QCD or asymptotic safety of gravity \cite{Hawking,Litim:2011cp}.
In general, the fixed point structure
of a given theory depends on its field content, the spacetime dimensionality,
the long-range or short-range nature of its interactions and the symmetries of
the action.

Scalar field theories with a  global $O(N)$ symmetry provide an important 
testing ground for fixed point studies. 
In three dimensions the $(\phi^2)^2$ theory displays a non-trivial IR fixed
point which determines the second-order phase transition between an $O(N)$
symmetric and the symmetry broken phase as realized in many physical  systems
ranging from   entangled polymers and water to ferromagnets or QCD with two
massless flavors of quarks \cite{ZinnJustin,Pelissetto:2000ek}.  The
$(\phi^2)^3$ theory also displays a line of first-order phase transitions whose
end point, in the limit of many scalar fields, qualifies as an UV fixed point 
\cite{Bardeen:1983rv,David:1984we}.
 
Supersymmetry represents the global symmetry which relates bosonic to
fermionic degrees of freedom. Supersymmetric theories are important 
candidates for extensions of the Standard Model. 
It is important to understand
how the fixed-point structure of a non-supersymmetric theory differs
from that of its supersymmetric extension, both in view of the IR and 
the UV behavior of the theory.

In this paper, we study fixed points of supersymmetric $O(N)$ models
which consist of an $N$-component scalar field coupled to $N$ Majorana 
fermions.  We employ non-perturbative renormalization group methods  a la
Wilson, based on the integrating-out of momentum modes from a path-integral
representation of the theory \cite{Polchinski:1983gv,Berges:2000ew,Bagnuls:2000ae}.
A particular strength of this continuum method is its flexiblity, allowing for the  
study of theories with strong correlations and large couplings. Furthermore, 
optimization techniques are available to control the physics content within 
systematic approximations \cite{Litim:2001up,Litim:2000ci,Litim:2001fd}.
In the past, this method has been successfully employed for the study of critical phenomena 
in a variety of settings including scalar theories, fermions, gauge theories and 
gravity \cite{Litim:1998yn,Litim:1998nf,Aoki:2000wm,Delamotte:2003dw,Pawlowski:2005xe,Gies:2006wv,
Reuter:2007rv,Sonoda:2007av,Rosten:2010vm,Litim:2010tt,Litim:2011cp}.  
It has recently been extended to include supersymmetric theories  
\cite{Vian:1998kv,Bonini:1998ec,Synatschke:2008pv,Gies:2009az, 
Synatschke:2009nm,Synatschke:2010jn,Synatschke:2010ub,Synatschke:2009da,Falkenberg:1998bg,Rosten:2008ih,Sonoda:2009df,Sonoda:2008dz}. 
Our prime interest here concerns the limit of many scalar fields $1/N\to 0$, 
where effects induced by the fields' anomalous dimensions are suppressed and 
a local potential approximation (LPA) becomes exact. Then full analytical 
fixed point results are obtained for the fixed points in the supersymmetric theory, 
allowing for a complete analytical understanding of the theory, analogous to the  
purely scalar theory \cite{Tetradis:1995br,Litim:1995ex,Litim:2002cf}.  

Supersymmetric $O(N)$ models have previously been investigated
with Dyson-Schwinger equations \cite{Dawson:2005uw} and 
with the large-$N$ expansion \cite{Bardeen:1984dx,Moshe:2003xn}. 
The three-dimensional theory has also been studied at finite 
temperatures, where supersymmetry is softly broken \cite{Moshe:2002ra,Feinberg:2005nx}.
The model has a peculiar phase structure concerning the breaking of the $O(N)$ symmetry: 
Additionally to the normal phases with a broken and an unbroken symmetry 
a phase with two $O(N)$ symmetric ground states and a phase 
with one symmetric and one non-symmetric ground state
have been found. In addition, there exists a supersymmetric analogue 
of the Bardeen-Moshe-Bander phenomenon 
\cite{Bardeen:1983rv}. The fate of this phenomenon at finite $N$ remains 
yet to be resolved \cite{Suzuki:1985uk,Gudmundsdottir:1984yk,Matsubara:1987iz}. 
   
The paper is organized as follows: First we introduce the
supersymmetric $O(N)$ model (Sec.~\ref{Susy}) and derive the non-perturbative
flow  equation for the superpotential in LPA (Sec.~\ref{RG}). We then
solve this equation analytically in the large-$N$ limit (Sec.~\ref{N}) and analyze 
the resulting fixed-point solutions (Sec.~\ref{FP}). We compute the   
universal scaling exponents  and compare our  results  with
those in the non-supersymmetric theory without fermions (Sec.~\ref{Uni}). 
We close with a discussion of our results (Sec.~\ref{Conclusion}).
Our conventions and a derivation of supersymmetric flow equations in
superspace is found in the appendix.

\section{Supersymmetry}\label{Susy}

In this section, we recall the definition of three-dimensional 
supersymmetric $O(N)$ models, which are built from $N$ real superfields
 \begin{equation}
 \Phi^{i}(x, \theta) = \phi^{i} + \bar{\theta}\psi^{i}(x) +
 \frac{1}{2}\bar{\theta}\theta F^{i}(x),\quad i=1,...,N.
 \label{eq:Superfield}
 \end{equation}
Each component of the superfield contains a real scalar field, 
a two-component Majorana spinor field and a real auxiliary field, 
$\Phi^i\sim(\phi^{i},\psi^{i},F^{i})$. We shall use a Majorana 
representation with imaginary $\gamma$-matrices  $\{\gamma^\mu\}
= \{\sigma_2,i\sigma_3,i\sigma_1\}$. Then the metric in  
$\{\gamma^{\mu}, \gamma^{\nu}\}= 2\eta^{\mu\nu}$ 
takes the form $\eta_{\mu\nu}=\hbox{diag}(1,-1,-1)$. 
A Majorana spinor is real in this representation and
$\bar{\psi}= (i\psi_2, \,-i\psi_1)$.  The supersymmetry variation
of the superfield is generated by the supercharge $\mathcal{Q}$ via  
$\delta_\epsilon\Phi^i=i\bar\epsilon\mathcal{Q}\Phi^i$, where the 
explicit form of the supercharge and further conventions are collected 
in appendix~\ref{sec:Conventions}. To construct a supersymmetric invariant
action we note that the $F$-term in the expansion (\ref{eq:Superfield}) 
transforms under supersymmetry transformations into a spacetime divergence
such that its spacetime integral is invariant.

In order to define an $O(N)$ symmetric, supersymmetric  action
we introduce the supercovariant derivatives
\begin{equation}
\cD = \frac{\partial}{\partial {\bar\theta}} + i\fdi\theta 
\quad \mbox{and} \quad \mathcal{\bar{D}} = -\frac{\partial}{\partial \theta} 
- i \bar\theta\fdi,
\label{eq:derivatives}
\end{equation}
which anticommute with the supercharges and thus map superfields into
superfields. Since the theory should be $O(N)$ invariant, the  superpotential
only depends on the invariant composite  superfield $R\equiv\frac{1}{2}\Phi^i\Phi_i$. 
In component form, it reads
\begin{equation}
R=\bar\varrho+(\bar{\theta}\psi_i)\phi^i+\frac{1}{2}\,\bar\theta\theta
\left(\phi^i F_i-\frac{1}{2}\bar{\psi}^i\psi_i\right),
\end{equation}
where the quantity $\bar\varrho\equiv\frac{1}{2}\phi^i\phi_i$ has been
introduced.  The starting point for further investigations will be the supersymmetric 
action
\begin{equation}
\mathcal{S} = \int d^3x\,\left[-\frac{1}{2}\Phi^{i}\bar{\cD}\cD\Phi_i + 2N\,W\left(\frac{R}{N}\right)\right]
\Big\vert_{\bar\theta\theta}
\label{eq:action}
\end{equation}
which contains a kinetic term
with supercovariant Laplacian $\bar{\cD}\cD$ as well as an  interaction  term,
given by the  superpotential $W$. We have already rescaled
the fields and the superpotential with $N$. An expansion in component fields
yields the Lagrangian density
\begin{eqnarray}
\mathcal{L}_{\rm off} &=& \frac{1}{2}\left(-\phi^i\Box\phi_i -i\bar{\psi}^i\fdi
\psi_i +F^2\right) +W^{\prime }\left(\frac{\bar\varrho}{N}\right)\phi_i F^i \nonumber\\
&-&
\frac{1}{2}W^{\prime}\left(\frac{\bar\varrho}{N}\right)\bar{\psi}^i\psi_i
-
W^{\prime\prime}\left(\frac{\bar\varrho}{N}\right)\frac{(\bar{\psi}^i\phi_i)(\psi^j\phi_j)}{2N},
\label{eq:offshell}
\end{eqnarray}
where  primes denote  derivatives with respect to $\bar\varrho/N$.
Eliminating the auxiliary field $F^i$ by its algebraic equation of motion,
$F^i=-W'(\bar\varrho/N)\phi^i$, yields the on-shell Lagrangian density
 \begin{eqnarray}
 \mathcal{L}_{\rm on}& =& 
 -\frac{1}{2}\phi^i\Box\phi_i 
 -\frac{i}{2}\bar{\psi}^i\fdi\psi_i 
 -\frac{1}{2}W'\left(\frac{\bar\varrho}{N}\right)\bar{\psi}^i\psi_i  \nonumber\\ 
 &&  - \bar\varrho\, W^{\prime 2}\left(\frac{\bar\varrho}{N}\right)
-W^{\prime\prime}\left(\frac{\bar\varrho}{N}\right)\frac{(\bar{\psi}^i\phi_i)(\psi^j\phi_j)}{2N}.
 \label{eq:onshell}
 \end{eqnarray}
From \eq{eq:onshell} we conclude that the potential for the bosonic field follows from the superpotential $W$ via
\begin{equation}\label{V_bosonic}
V(\bar\varrho)=\bar\varrho\, W^{\prime 2}\left(\frac{\bar\varrho}{N}\right).
\end{equation}
Note that for a polynomial superpotential $W(\bar\varrho/N)$ which for 
large $\bar\varrho$ tends to $W\sim \bar\varrho^{\,n}$ 
we do \textit{not}
expect supersymmetry breaking in our non-perturbative  renormalization group studies.

\section{Renormalization group}\label{RG}

\subsection{Supersymmetric flows}
In order to analyze the phase transition and the low-energy behavior of 
supersymmetric sigma models we resort to Wilsonian renormalization group techniques. 
Specifically, we adopt  the framework of the effective average action based on the infinitesimal integrating-out of degrees of freedom with momenta $q^2$ larger than some infrared momentum scale $k^2$.
 In consequence, the effective action becomes a scale-dependent effective action
 $\Gamma_k$ which interpolates between the microscopic action $S$ in the UV and the full quantum effective action in the IR, where $k\to 0$.
The scale-dependence of $\Gamma_k$ is given by an exact functional 
differential equation~\cite{Wetterich:1992yh} 
\begin{equation}
\partial_t\Gamma_k=\frac{1}{2}\Str \left\{\partial_t
R_k\left(\Gamma^{(2)}_k+R_k \right)^{-1} \right\},  \label{eq:LPA1}
\end{equation}
where $t=\ln(k/\Lambda)$. The function $R_k(q^2)$ denotes the momentum cutoff.
It  obeys $R_k(q^2)\to 0$ for $k^2/q^2\to 0$, $R_k(q^2)>0$ for $q^2/k^2\to 0$  and
$R_k(q^2)\to\infty$ for $k\to\Lambda \rightarrow \infty$,
where $k=\Lambda$ stands for the initial scale in the UV. The stability and convergence 
of the RG flow \eq{eq:LPA1} is controlled through adapted, optimized choices of the momentum cutoff 
\cite{Litim:2001up,Litim:2001dt,Litim:2002cf}. Furthermore,
$\Gamma^{(2)}_k$ denotes  the second functional derivative of $\Gamma_k$ with
respect to the fields
according to
\begin{equation}
  \left(\Gamma_k^{(2)}\right)_{ab}=\frac{\overrightarrow{\delta}}{\delta\Psi^a}
\Gamma_k\frac {\overleftarrow{\delta}}{\delta\Psi^b},
\end{equation}
where the indices $a,b$ summarize field components, internal and Lorentz  
indices as well as coordinates. Note that $\Psi$ is merely a collection of fields and not a superfield

Following the construction in \cite{Synatschke:2008pv,Gies:2009az,Synatschke:2009nm,Synatschke:2010jn,Synatschke:2010ub},
it is essential that the regulator term $\Delta S_k$ preserves  
both the $O(N)$-symmetry and supersymmetry of the classical 
theory. Being quadratic in the fields it should be the superspace 
integral of $\Phi^i R_k(\bar\cD\cD)\delta_{ij}\Phi^j$. 
Using the anticommutation relation $\{\cD_k,\bar{\cD}_l\}=-2(\gamma^{\mu})_{kl}\partial_{\mu}$ 
for the supercovariant derivatives, we have
\begin{equation}
 \Big(\ha\bar\cD\cD\Big)^{2n}=(-\Box)^n,
\end{equation}
such that a supersymmetric and $O(N)$-invariant regulator term
is the superspace integral of
\begin{equation}
\Phi_i
R_k(\bar\cD\cD)\Phi^i=\Phi_i\left(r_1(-\Box)-r_2(-\Box)\frac{\bar\cD\cD}{2}\right)\Phi^i.
\label{eq:regulator}
\end{equation}
Expressed in component fields, we find
\begin{equation} 
\Delta S_k=\ha\int \,(\phi,F) R^{\rm B}_k\, {\phi \choose F}
+\ha\int \, \psib R^{\rm F}_k\psi\,.\label{eq:lpa11}
\end{equation}
In momentum space $i\partial_\mu$ is replaced
by $p_\mu$ and the bosonic  and fermionic  momentum cutoffs $R^B_k$ and
$R^F_k$ respectively are of the form
\begin{eqnarray}
R_k^{\rm B}&=&\begin{pmatrix}
      p^2r_2&r_1\\r_1&r_2
      \end{pmatrix}\otimes\mathds{1}_{N}
      \nonumber\\
R_k^{\rm F}&=&-\left(r_1+r_2\slashed{p}\right)\otimes\mathds{1}_{N}\,.
\end{eqnarray}
Note that the requirements of manifest supersymmetry imposes a link between the
bosonic and fermionic momentum cutoffs,
leaving two free functions $r_1\equiv r_1(p^2/k^2)$ and $r_2\equiv r_2(p^2/k^2)$ 
at our disposal. Such supersymmetric cutoffs have been introduced
for the $N=1$ model in two and three dimensions in \cite{Gies:2009az,Synatschke:2010ub}.

There exist no Majorana fermions in three Euclidean spacetime dimensions. With
respect to  the supersymmetric $O(N)$ model we could thus
analytically continue the flow equation in Minkowski spacetime to imaginary time or alternatively just ignore the fact that the Majorana condition 
is not compatible with Lorentz invariance in Euclidean spacetime 
\cite{Moshe:2002ra}. Both approaches lead to identical flow equations in Euclidean 
spacetime, cf. \cite{Synatschke:2010ub}. 

\subsection{Local potential approximation}

 Next we turn to the supersymmetric RG flow in the local potential approximation. Here, one keeps the 
leading order term in a superderivative expansion such that the effective action 
(with Lorentzian signature) reads
\begin{align}
\Gamma_k[\Phi] =& \int\! d^3x\, 
\left[-\frac{1}{2} \Phi^i \bar\cD\cD \Phi_i + 2 N\,W_k\left(\frac{R}{N}\right)
\right]\Big|_{\bar{\theta}\theta}
\nonumber\\
=&
\ha\int d^3x\,\left(\partial_\mu\phi^i\partial^\mu \phi_i -i\bar{\psi}^i\fdi
\psi_i +F^2\right)
\label{eq:LPA3}
\\ 
\nonumber
+&\int d^3x\left(W_k^{\prime }
\frac{2\phi^i F_i -\bar{\psi}^i\psi_i}{2}
 -
\frac{W_k^{\prime\prime}
(\bar{\psi}^i\phi_i)(\psi^j\phi_j)}{2N}\right),
\end{align}
where the prime denotes the derivative with respect to $\bar\varrho/N$.
The flow of the renormalized superpotential $W_k(\frac{\bar\varrho}{N})$
in Euclidean space  is obtained  by projecting the flow (\ref{eq:LPA1})
onto the term linear in the auxiliary field $F$ and performing a Wick
rotation (see appendix~\ref{sec:Derivation} for its derivation in superspace).
The function $r_1$ acts as IR regulator but not as UV-regulator,
in contrast to $r_2$ which serves both as IR and UV regulator. Thus we
use $r_1$ as regulator in what follows\footnote{In preliminary 
studies we did include the regulator $r_1$ and got
almost identical results.}. Then we find
  \begin{equation}
    \begin{split}
  \partial_t W_k =
 -\frac{1}{2}\int\!\!\frac{d^3p}{(2\pi)^3}\,\partial_tr_2
 \left(\frac{N-1}{N}\frac{W_k^{\prime}}{(1+r_2)^2p^2 + W_k^{\prime 2}}\right.\\ 
 \left.+\frac1N \frac{W_k^{\prime} +
 2(\bar\varrho/N) W_k^{\prime\prime}}{(1+r_2)^2p^2 + (W_k^{\prime} +
 2(\bar\varrho/N) W_k^{\prime\prime})^2}\right)\label{eq:LPA5}.
 \end{split}
   \end{equation}   
Similar to the bosonic $O(N)$ model the flow receives
contributions from the $N-1$ Goldstone modes (the first term) and 
from the radial mode (second term).

Next we specify the function $r_2(p^2/k^2)$.
Following \cite{Litim:2001up,Litim:2000ci,Litim:2001fd,Synatschke:2010ub}  
we choose the  optimized regulator function 
\begin{equation}
 r_2(p^2) = \left(\frac{k}{|p|}-1\right)\theta (k^2-p^2).\label{eq:LPA7}
   \end{equation}
This choice implies $\partial_t r_2$ to vanish identically for $p^2>k^2$, and
the inverse propagators
\begin{equation}
(1+r_2)^2\,p^2+X=\left\{
\begin{array}{ll}
      p^2+X&{\rm for}\ p^2>k^2\\
      k^2+X&{\rm for}\ p^2<k^2 
\end{array}
\right.
\nonumber
\end{equation}
become flat (momentum independent) in the regime where the right-hand 
side of \eq{eq:LPA5} is non-vanishing. In the LPA this is a solution to the 
general optimization condition of \cite{Litim:2000ci,Litim:2001up,Litim:2001fd} 
and is therefore expected to lead to improved convergence and stability of the RG flow.
Equally important,  the momentum integrals in (\ref{eq:LPA5}) can be performed
analytically, leading to 
\begin{eqnarray}
\partial_kW_k
&=&-\frac{k^2}{8\pi^2}\left(1-\frac{1}{N}\right)\frac{W_k^{\prime}}{k^2 + W_k^{\prime 2}}
\nonumber\\
&& -\frac{k^2}{8\pi^2}\frac1N \frac{W_k' + 2(\bar\varrho/N) W_k''}{k^2 + (W_k' + 2(\bar\varrho/N) W_k'')^2}\,.
\label{eq:LPA9}
\end{eqnarray}
With given initial condition $W_{k=\Lambda}(\bar\varrho/N)\equiv W(\bar\varrho/N)$ 
at the UV scale $\Lambda$ this flow equation uniquely determines the superpotential in the
infrared limit $k \rightarrow 0$.  For $N=1$ it reduces to the three-dimensional Wess-Zumino model studied in \cite{Synatschke:2010ub}.

In order to write the flow equation \eq{eq:LPA9} in a scale-invariant form it is convenient to define a dimensionless field variable $\rho$ as well as a dimensionless superpotential $w$ and a dimensionless scalar potential $v$. The canonical mass dimension of the fields and potentials are $[\bar\varrho]={d-2}$, $[V]=d$ and $[W]=d-1$ in $d$ spacetime dimensions. 
We therefore introduce the dimensionless quantities
\begin{eqnarray}
\rho&=&\frac{8\pi^2}{N}\frac{\bar\varrho}{k}\,\quad\mbox{and} \quad
w(\rho)=8\pi^2\,\frac{W(\frac{\bar\varrho}{N})}{k^2}.
\label{eq:LPA11}
\end{eqnarray}
Note that we have also rescaled an irrelevant numerical factor into the potential and the fields. It is understood that $w$ is also a function of the RG scale parameter, though this is not spelled out explicitly. Similarly, we define the dimensionless bosonic potential $v$ as
\begin{equation} \label{v}
  v(\rho)=
  \frac{8\pi^2}{N}\,\frac{\bar\varrho}{k}\,\left(\frac{W'(\frac{\bar\varrho}{N})}{k}\right)^2\equiv\rho\,w'^2(\rho),
\end{equation} 
where \eq{V_bosonic} and \eq {eq:LPA11} have been used.
Thus, by substituting (\ref{eq:LPA11}) into (\ref{eq:LPA9}) we
end up with the following flow equation for the dimensionless
superpotential,
\begin{equation}
\partial_t w-\rho w'+2w
=-\frac{(1-\frac{1}{N})w^{\prime}}{1 + w'^2} -\frac{\frac1N (w' + 2\rho w'')}
{1 + (w' + 2\rho w'')^2}.\label{eq:LPA13}
\end{equation}

\subsection{Large-$N$ limit}\label{N}

In the large-$N$ limit the Goldstone modes fully dominate the
dynamics and the contribution of the radial mode becomes a subleading
effect. It follows that the anomalous dimension of the 
Goldstone modes vanish, as no momentum-dependent 
two-point function exists that contribute to the running of the kinetic 
term of these modes to leading order in $N$. This is a particular
feature of the bosonic $O(N)$ models \cite{ZinnJustin}  and their
supersymmetric extensions\footnote{For example,
Yukawa-type systems may have large anomalous dimensions
in the large-$N$ limit \cite{Braun:2010tt}.}.
Consequently, the LPA approximation becomes exact
for $N\to\infty$.

In this limit, the RG equation for the first derivative of 
the superpotential $u(\rho)\equiv w'(\rho)$ becomes
\begin{equation}\label{eq:LPA15}
\partial_t u+\partial_\rho u \left[1-\rho-u^2\,f(u^2)\right]=-u
\end{equation}
with 
$ f(x)=(3+x)/(1+x)^2$. We note that the second-order partial 
differential equation \eq{eq:LPA13} has turned into a 
first-order one in this limit, which is solved analytically with the {method of
characteristics}. The first characteristic reads $ue^t=$ const. and the second one is 
\begin{equation}
\frac{\rho-1}{u}- F(u)={\rm const.}
\label{eq:LPA17}
\end{equation}
with
\begin{equation}
F(u)=\frac{u}{1+u^2}+2\arctan u
\label{F}
\end{equation}
and $F'(u)=f(u^2)$. Altogether, we find
\begin{equation}
\frac{\rho-1}{u}-F(u)=G(ue^t)
\label{eq:LPA21}
\end{equation}
for all $\rho\ge0$, where the function $G(ue^t)$ is determined by the boundary
conditions for $u(\rho)$, imposed at the initial UV scale $k=\Lambda$. The
validity of the solution \eqref{eq:LPA21} is confirmed by direct insertion into
(\ref{eq:LPA15}).
For completeness, we also give the RG equation for the bosonic potential. 
Using \eq{v} and \eq{eq:LPA17} we obtain
\begin{equation}\label{dtv}
\partial_t v+3 v-\rho\,v'=(v-\rho\,v')\frac{\rho-v}{(\rho+v)^2}.
\end{equation}
In passing we note that up to minor modifications eq.~\eqref{eq:LPA15} holds
for general spacetime dimensions away from $d=3$. The canonical
mass dimension of $u$ is one for all dimensions and  the 
dependence on spacetime dimensionality, therefore, 
only enters via the field variable leading to the replacement 
of $(-\rho)$ by $(2-d)\rho$ in \eq{eq:LPA15}. 
This modifies the second characteristic equation whose solution is 
expressed in terms of the  hypergeometric function for arbitrary 
dimension $d\neq1$. Below, we restrict ourselves to the case $d=3$.

\begin{figure}[t]
\begin{center}
{\includegraphics[width=\columnwidth]{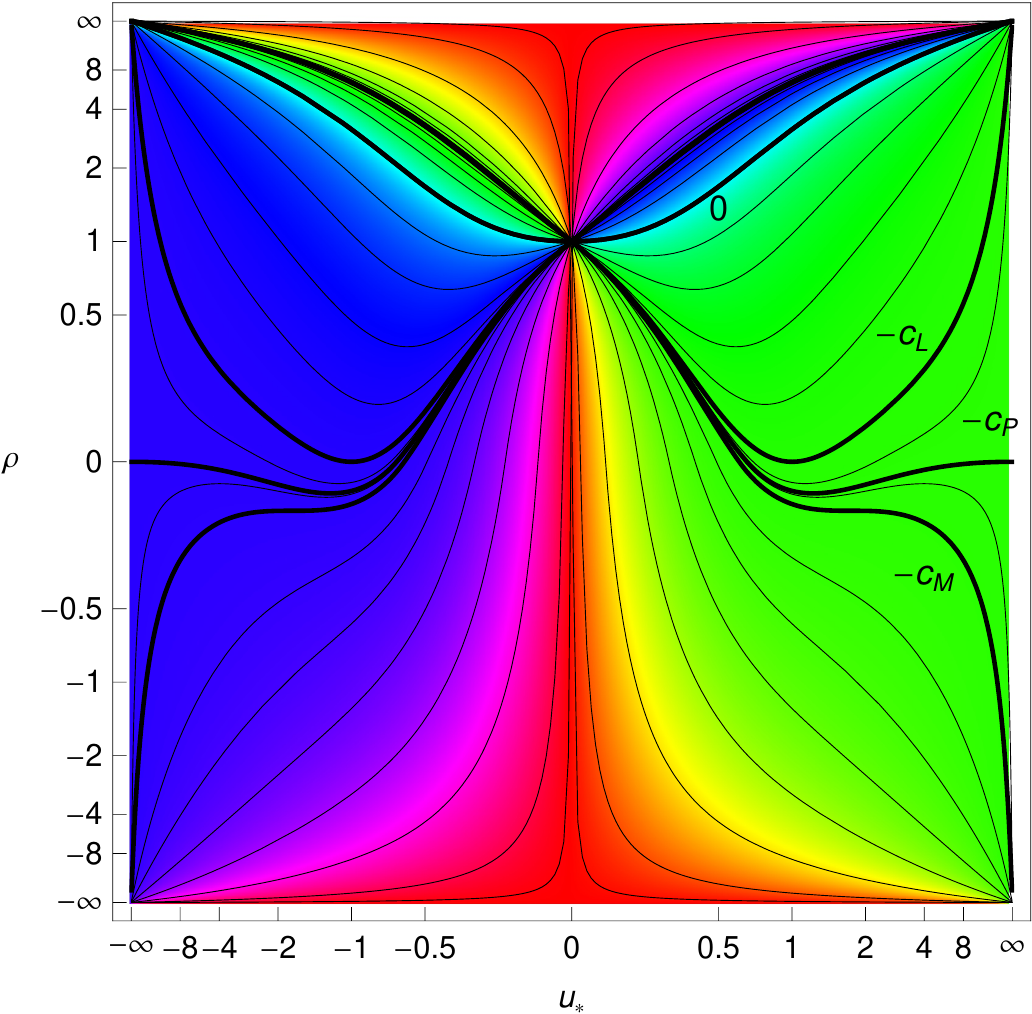}}\\
\vspace{.5\baselineskip}
{\includegraphics[width=\columnwidth]{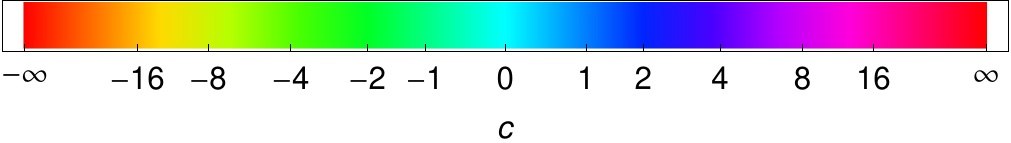}} 
\caption{Supersymmetric fixed point solutions $\rho(u_*)$ for all fields $\rho$
and all superfield potentials $u_*$, color-coded by the free parameter $c$ (both
axes are rescaled as $x\to\frac{x}{1+|x|}$ for display purposes). Thin lines are
included to guide the eye, thick lines correspond to distinguished values for $c$ $(|c|=0,c_L,c_P,c_M)$ as defined in main text.}
\label{fig:FPColorScheme} 
  \end{center}
\end{figure}

\begin{figure*}[t] 
\centering{
\includegraphics[width=.48\textwidth]{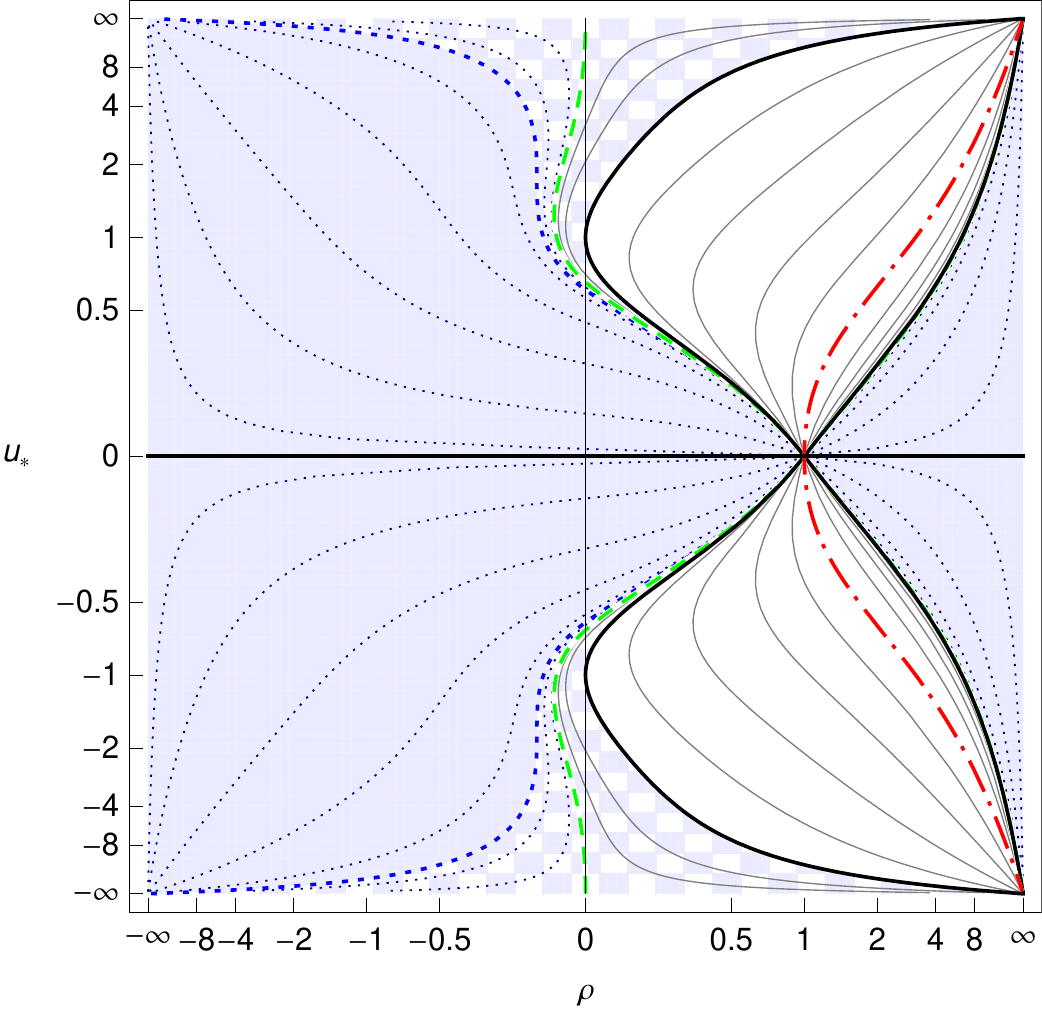}\hfill
\includegraphics[width=.48\textwidth]{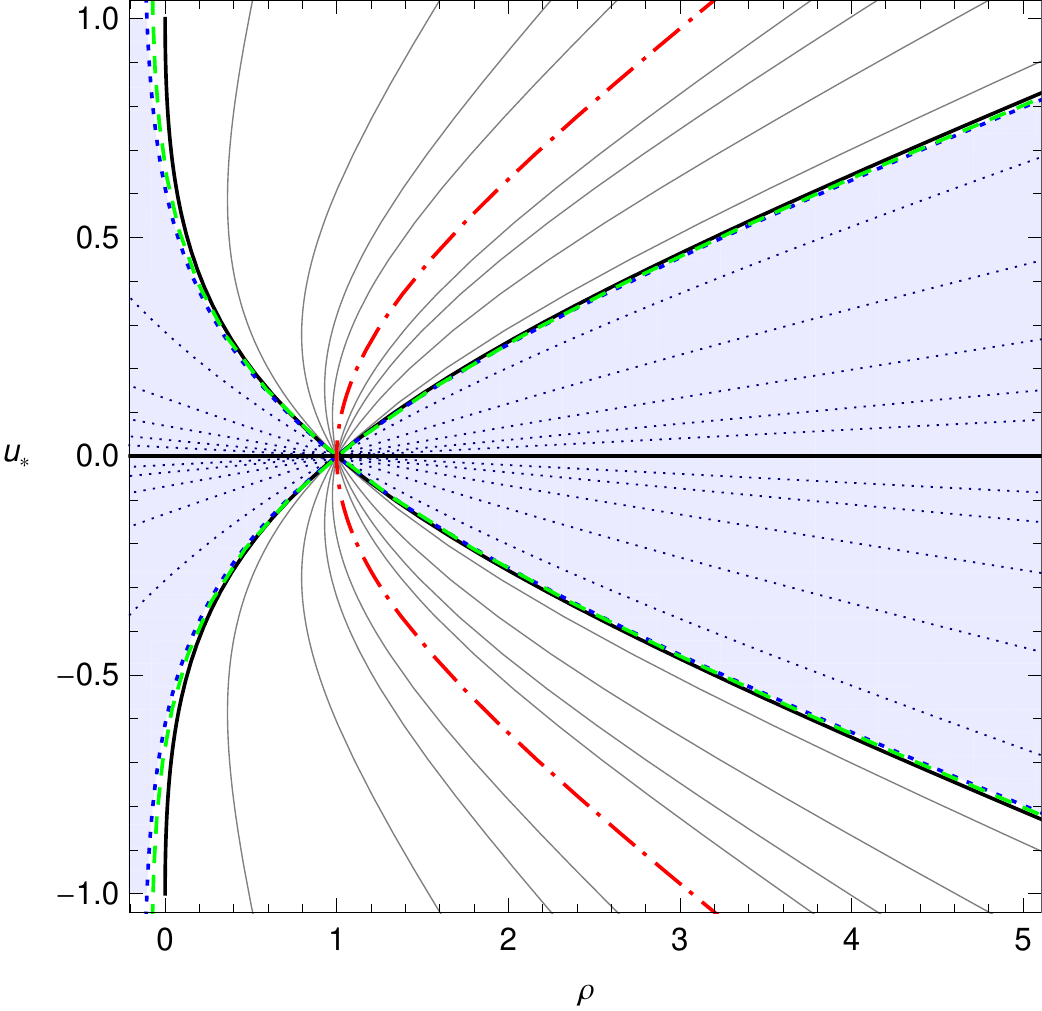}   
}
\caption{\label{fig:FPfull}Supersymmetric fixed point solutions $u_*(\rho)$
according to \eq{eq:FP1}, covering the entire parameter range for $c$. With
decreasing $c$, fixed point curves rotate counter-clockwise around
$(\rho,\uas)=(1,0)$ starting with  $c=\infty$ where $\uas=0$ (horizontal line),
passing through $c=0$ (red, dashed-dotted line), completing a rotation of {180\textdegree} at $c=-\infty$ (horizontal
line). Further special lines refer to $|c|=c_M$ (blue dashed), 
$|c|=c_P$ (green, long dashed),
$|c|=c_L$ (black, full lines), see main text. \emph{Left panel:} 
fixed point solutions for all fields (both axes are rescaled as
$x\to\frac{x}{1+|x|}$ for display purposes). \emph{Right panel:} fixed point solutions
for physical fields in the vicinity of $\rho=1$.  
}
\end{figure*}

\section{Fixed points}\label{FP}

\subsection{Supersymmetric fixed points}

Fixed points are the scale-independent solutions of \eq{eq:LPA15}, i.\,e.
solutions satisfying $\partial_t u=0$. Besides the  Gaussian fixed-point
solution $u_*\equiv 0$, non-trivial fixed points follow from \eq{eq:LPA21} in the limit where $G(ue^t)$ becomes a $t$-independent constant. 
The classification of solutions of
 \begin{equation}
\rho = 1 + H(u_\ast)+ c\,u_\ast,\quad H(\uas)=\uas\,F(\uas), \label{eq:FP1}
 \end{equation}
 where $F(\uas)$ is given by \eq{F},
then depends only on the real parameter $c$.
With $|\uas|\in[0,\infty)$ and for a fixed $c$ \eq{eq:FP1} identifies 
the range of achievable field values. Candidates for
{physical} fixed points $\uas(\rho)$ are those solutions which extend over all
fields $\rho\in [0,\infty)$.   Fig.~\ref{fig:FPColorScheme} and
Fig.~\ref{fig:FPfull}  display the entire set of solutions to \eq{eq:FP1} for all $c$. 
Note that Fig.~\ref{fig:FPColorScheme} shows the function $\rho(\uas)$ whereas 
the relation $\uas(\rho)$ is displayed  in Fig.~\ref{fig:FPfull}.

The space of solutions enjoys some internal symmetry. Since $H(u_*)$ is an even
function, solutions only depend on the absolute value of $c$, i.\,e. any
solution $u_*(\rho)$ with parameter $c$ is equivalent to the
reflected solution $-u_*(\rho)$ with parameter $-c$.
Both solutions lead to identical scalar potentials $v_\ast$ and therefore we
may restrict our discussion to $c\ge 0$.

We now discuss \eq{eq:FP1} in more detail. All curves pass through
\begin{equation}\label{zero}
(\rho,u_\ast)=(1,0)
\end{equation}
which follows immediately from \eq{eq:FP1} due to  $H(0)=0$. 
As can be seen from Fig.~\ref{fig:FPfull} the fixed point
solutions fall into two distinct classes, and
solutions in the same class show the same global behavior.
Depending on the value of $c$ the solution $u_\ast$ is either defined for all 
real $\rho$ or it has a turning point at $\abs{\rho_s}<\infty$ and is 
only defined for $\rho\in[\rho_s,\infty)$. In the latter case the solution 
has two branches bifurcating at $\rho=\rho_s$.  The value of $\rho_s$ will 
be determined below.

Next we discuss some limiting cases of interest. For small $\uas$
we conclude from \eq{eq:FP1}  that 
\begin{equation}\label{small}
\rho-1=c\,\uas+3\,\uas^2+{\cal O}(\uas^4).
\end{equation} 
Hence, the potential is analytical in $\rho-1$ in the vicinity of 
$\rho=1$ for all $c$, except for $c=0$ where it becomes non-analytical 
with $\uas\propto \sqrt{\rho-1}$. 
Eq. \eq{small} implies that all fixed-point solutions have 
one simple zero at $\rho=1$ with finite $\uas'(1)$ except for
$c=0$ where $\uas'(1)$ diverges. Consequently, the scalar fixed-point 
potentials $v_*=\rho u^2_\ast$  possess two minima at
 \begin{equation}
\rho=0\quad\hbox{and}\quad \rho = 1\,,
 \label{kappastern}
 \end{equation}
the first one being a simple zero. The second minimum is a double
zero for $c\neq 0$ and a simple zero for $c=0$.

 \begin{figure*}[t]
\includegraphics[height=8.2cm]{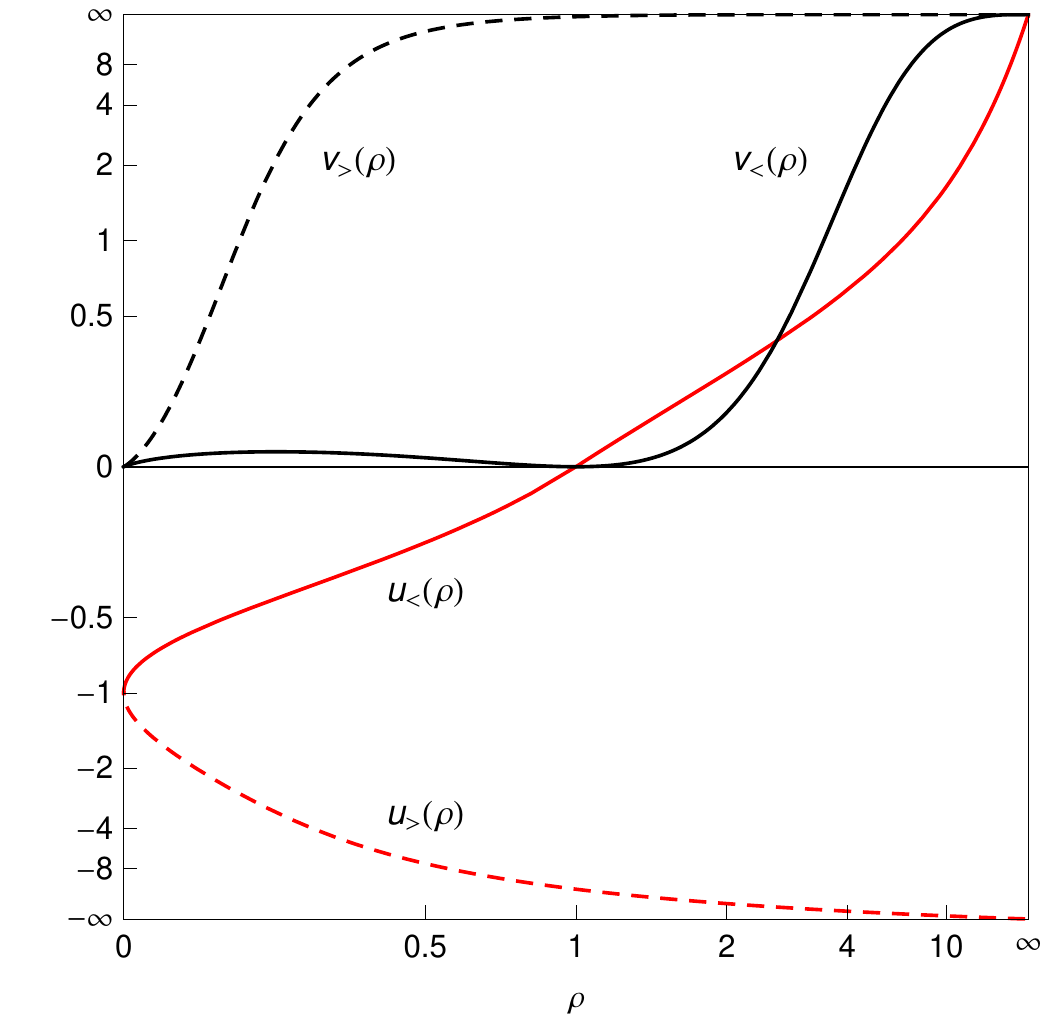}\hfill
\includegraphics[height=8.2cm]{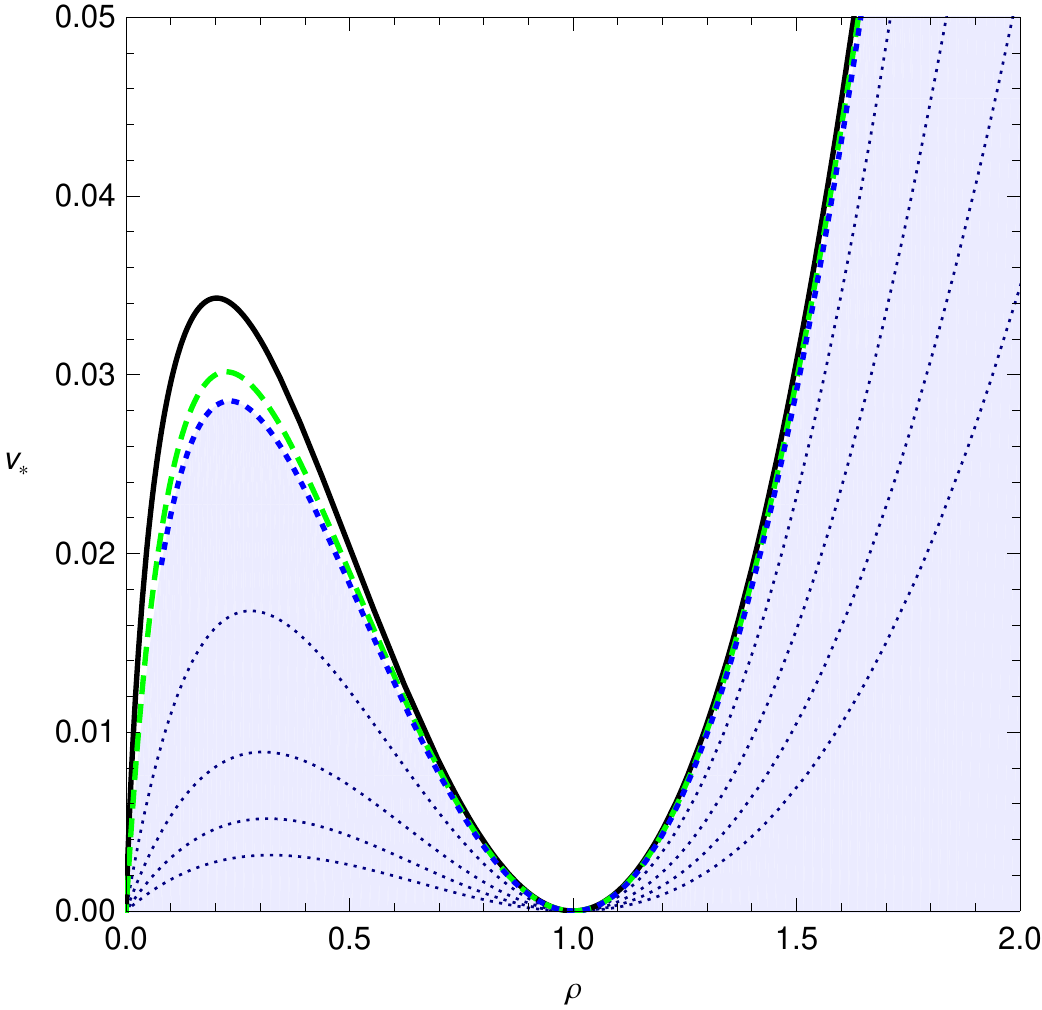} 
\caption{\emph{Left panel:} Fixed point solutions $u_\ast$  and 
fixed point potentials $v_\ast=\rho\,\uas^2$ at $|c|=c_L$ showing the 
two branches $v_<, u_<$ (full lines) and $v_>, u_>$ (dashed lines).
\emph{Right panel:} The scalar fixed point 
potential $v_\ast(\rho)$ as a function of the parameter 
$c$ with $c_{L}$ (black line), $c_P$ (green, long dashed), $c_M$ 
(blue, short dashed) and $c=a^n\,c_L$, ${a=2^{1/4}}$ 
with $n=1.0, 2.3, 3.6, 4.9$ (blue, dotted). For $c_L$ and $c_P$ 
just one branch is plotted.}
\label{fig:FPEbene3}
\end{figure*}

In the large-$\uas$ limit of \eq{eq:FP1}, we find
\begin{equation}\label{largeR}
\rho=\pi\,|\uas|+c\,\uas+{\cal O}(1/\uas^2)\,.
\end{equation}
Thus the asymptotic behavior of $\uas$ is given by
\begin{equation}\label{large|\uas|}
  \begin{split}
\uas=&\frac{\rho}{c+\pi}+{\rm subleading}\quad (\uas>0)\,,\\
\uas=&\frac{\rho}{c-\pi}+{\rm subleading}\quad (\uas<0)\,.
\end{split}
\end{equation}
If $\abs{c}> c_P$, with 
\begin{equation}\label{cP}
c_P=\pi\,,
\end{equation} the expansions extend towards $\rho\to\pm\infty$, respectively.
Together with the boundedness of $H(\uas)$ we conclude that $\uas(\rho)$
is defined for all real $\rho$. The expansions
correspond to asymptotically large fields $\rho\gg 1$   in the physical regime. 
At $|c|=c_P$ the leading term in \eqref{largeR} vanishes and, depending on 
the sign of $c$, one of the asymptotic solutions is replaced 
by $\uas\sim\rho^{-1/2}$ thus corresponding to a small field regime $\rho\ll 1$.
For $|c|< c_P$ both expansions extend towards $\rho\to+\infty$. We
conclude that $\uas$ has, simultaneously, two 
asymptotic expansions for large positive $\rho$. This implies that
$v_\ast$  displays a loop consisting of two branches $v_<$ and $v_>$ which 
coincide at $\rho=\infty$ and possibly at some $\rho=\rho_s<\infty$
where $u_\ast$ has infinite slope. The latter
condition determines the turning point $\rho_s$ as the simultaneous solution of
\begin{equation}\label{rho_s}
\rho_s=\frac{1-u_s^2}{(1+u_s^2)^2}
\end{equation}
together  with \eq{eq:FP1}, leading to
\begin{equation}\label{cL}
|c|=\frac{1}{|u_s|}\left(\frac{u_s^2(3+u_s^2)}{(1+u_s^2)^2}+H(u_s)\right)\;,
\end{equation}
where $u_s\equiv \uas(\rho_s)$. The degenerate solutions extend over the 
whole physical regime $\rho\geq 0$, provided that $\rho_s\le 0$. From \eq{rho_s} 
it follows that the equal sign holds for $u_s^2=1$ leading with \eq{cL} to $|c|=c_L$, 
where
\begin{equation}\label{cL2}
c_L=\frac12(\pi+3)\approx 3.071\,.
\end{equation}
For $|c|=c_L$, both $u_<$ and $u_>$ have infinite slope at vanishing field 
with the non-analytical behavior 
\begin{equation}
\label{BMB}
\frac{d\uas}{d\rho}=\pm\frac{1}{\sqrt{\rho}}+{\rm subleading}
\end{equation}
and $u_s=\mp1$ for $c=\pm c_L$ (see Fig.~\ref{fig:FPEbene3}, left panel). In contrast,
for $c_L< |c|< c_P$, the behavior at vanishing field is analytic.
The turning point \eq{rho_s} exists for small $0\le |c|\le c_M$ as long as $d^2\rho/d\uas^2|_{\rho_s}$ does not vanish, which happens at $u_s^2=3$ 
leading with \eq{cL} to $|c|=c_M$, where
\begin{equation}
c_M=2\left(\frac\pi3+\frac{5\sqrt{3}}{16}\right)\approx 3.177\,.
\label{cM}
\end{equation}
We note that $c_P<c_M$ and conclude that the fixed point solutions in 
the parameter regime $c_P\le |c|< c_M$ are \emph{single-valued in the 
physical regime} but multi-valued in the non-physical regime $\rho<0$. 
For all $c_M\le |c|$, fixed point solutions are single-valued on 
the entire real axis.  In Fig.~\ref{fig:FPEbene3}, right panel the 
scalar fixed point potential $v_\ast$ for different values of $c$ is displayed.

\subsection{Exactly marginal coupling} 
Next we discuss the physical meaning of the parameter $c$.
To this end we employ the polynomial expansion of the RG-time 
dependent superpotential $u(t,\rho)$ which satisfies the
flow equation \eqref{eq:LPA15}. For a typical initial
condition $u_\Lambda=\tau_1(\rho-\rho_0)$  there always exists a 
node $\rho_0(t)$ around which we can perform a Taylor expansion:
\begin{equation}\label{expansion}
u(t,\rho)=\sum_{n=1}\frac{1}{n!}\tau_n(t)\left(\rho-\rho_0(t)\right)^n\,. 
\end{equation}
Inserting this ansatz into the flow equation \eqref{eq:LPA15} 
we read off the flow equations for $\rho_0$ and the couplings
$\tau_n$ entering the Taylor expansion
\begin{eqnarray}
\label{vev}
\partial_t\rho_0&=&1-\rho_0\\ 
\label{marginal}
\partial_t \tau_1 &\equiv&0\\ 
\partial_t \tau_2 &=&6\tau_1^3+\tau_2
\end{eqnarray}
and similarly to higher order. Several comments are in order
at this point. Firstly, the running of the vev $\rho_0(t)$ 
is independent of all the other local couplings. This property is typical for a supersymmetric flow 
and has previously been observed in \cite{Synatschke:2009nm,Synatschke:2010ub}. 
The fixed point is obtained for $\rho_0=1$. Secondly, the system 
of algebraic equations describing the $t$-independent fixed point couplings 
can be solved recursively. This  leads to fixed point 
couplings $\tau_n(\tau_1)$ for all $n\ge 2$ as functions 
of  $\tau_1$.  Inserting \eq{expansion} into the expansion of the scalar field
potential $v=\rho \,u^2=\sum_{n=2}\lambda_n/n!\,(\rho-\rho_0)^n$ and evaluating it on the fixed point
leads to the fixed point values
\begin{eqnarray} 
\lambda_{2}&=&2\,\tau_1^2\\
\lambda_3&=&6\,\tau_1^2\,(1-6\tau_1^2)  \\
\lambda_3&=&-24\tau_1^4(1-45\tau_1^2)  
\end{eqnarray}
and similarly to higher order.
Clearly, the {weak (strong)} coupling regimes correspond to small (large) $\lambda_2$ 
and hence small (large) $\tau_1$ respectively. 
Also,  on the level of the scalar field potential the critical behavior is 
independent of the sign of $\tau_1$. Finally, and most importantly, the
coupling $\tau_1$ remains un-renormalized under the supersymmetric RG flow
\eq{marginal}. 
Therefore $\tau_1$ corresponds to an \emph{exactly marginal coupling}, and fixed
points can be classified according to the value of the linear (dimensionless) 
superfield interaction $\tau_1$ which relates to the free parameter $c$ in the 
analytical solution \eq{eq:FP1} as
\begin{equation}\label{cLambda}
c=\frac{1}{\tau_1}\,.
\end{equation}
This relation can be shown by inserting expansion \eqref{expansion} 
into the fixed point equation \eqref{eq:FP1}.
The presence of the exactly marginal coupling $\tau_1$ explains the existence 
of a line of fixed points.

\subsection{Line of fixed points}
In summary, the following picture has emerged. Fixed point solutions are characterized 
by the dimensionless linear superfield coupling $\tau_1=1/c$ in the vicinity of 
the node $\rho_0\neq 0$. In the \emph{weakly coupled regime}
\begin{equation}
c_P\le |c|
\end{equation}
a unique  fixed point solution exists covering the whole physical domain 
$\rho\geq 0$. This includes  the Gaussian fixed point $\tau_1=0$.  
In the \emph{ intermediate coupling regime}
\begin{equation}
c_L\le |c|< c_P
\end{equation}
two separate fixed point solutions $u_<$ and $u_>$ exist. The former
solution has a node at $\rho_0=1$ whereas the other solution has no
node,  see Fig.~\ref{fig:FPEbene3}, 
left panel. Therefore, the corresponding scalar field potentials $v_<$ ($v_>$) 
have two minima at \eq{kappastern} (one minimum at $\rho=0$). Both are 
analytical functions of $\rho$ in the vicinity of their global minima.  For $|c|=c_L$, 
the potential becomes non-analytical for either of them at $\rho=0$ in a manner 
reminiscent of the Bardeen-Moshe-Bander phenomenon in the purely scalar theory 
\cite{Bardeen:1983rv}.
In the \emph{strong coupling regime}
\begin{equation}
|c| < c_L
\end{equation}
the theory becomes so strongly coupled that $du/d\rho|_{\rho_s}$ diverges in the 
physical regime, and hence no fixed point solution exists which extends over all 
fields. Therefore, the supersymmetric $O(N)$ model displays  a line of fixed points
 which bifurcates at $|c|=c_P$ into two fixed points, and then terminates at $|c|=c_L$. 
 
Finally we note that the solution with $c=0$ is closely linked 
to the Wilson-Fisher fixed point in the purely bosonic 
model \cite{Litim:1995ex,Tetradis:1995br,Litim:2002cf}.
The precise relation is discussed in Sec.~\ref{WFFP} below.

\section{Universality}\label{Uni}

\subsection{Critical exponents}\label{CE}
Fixed point solutions are characterized by universal critical scaling exponents. 
The exponents can be deduced from the RG equations in several ways. Within a 
polynomial approximation up to order $n$, we expand 
$u(\rho)=\sum_{i=1}^n b_i (\rho-b_0)^i/i!$ in terms of the $n+1$ 
couplings $b_i(t)$. From their beta-functions $\beta_i\equiv \partial_tb_i$, 
the universal exponents follow as the negative of the 
 eigenvalues $\theta^I$  of their stability matrix 
$B_i^{\ j}=\partial\beta_i/\partial b_j |_{b=b^\ast}$ as 
$B \,v^I=-\theta^I\,v^I$ with eigenvectors $v^I.$
Using the flow equation, we find
\begin{align}\label{theta1}
	\theta=-1,0,1,2,3,\cdots
\end{align}
both numerically and analytically. In fact, the LPA approximation has become 
exact in the large-$N$ limit, and hence the correct scaling exponents are achieved 
to every order in the polynomial approximation. We note that this analysis relies 
on local information of the RG flow in the vicinity of $u=0$, showing that the 
scaling \eq{theta1}  is achieved mathematically for all $0<|c|<\infty$. 
Physically, however, the analysis is not sensitive to the global behavior of 
the solution, and consequently cannot detect that $|c|=c_L$ denotes a physical 
endpoint. Also, the case $c=0$ requires special care as an analytical expansion 
about $u=0$ is no longer applicable.

\subsection{Eigenperturbations}
Interestingly, the critical exponents and eigenperturbations can also be calculated 
analytically without resorting to a polynomial expansion. 
To that end, we consider small fluctuations $\delta u$ about the fixed-point 
superpotential such that
$u(t,\rho)=u_*(\rho) +\delta u(t,\rho)$. Linearizing the flow equation in
$\delta u$ leads to the fluctuation equation
\begin{equation}\label{flowdelta}
\partial_t\,\delta u=
\frac{u_*}{u_*'}\left(\partial_\rho-\frac{(u_*u_*')'}{u_*u_*'}\right)\,\delta u,
\end{equation}
where  primes denote a derivative with respect to the function's argument.
Since the right-hand side is independent of $t$, the differential equation 
\eq{flowdelta} can be factorized  via separation of variables 
$\delta u(t,\rho)=f(t)g(\rho)$ with
\begin{eqnarray}
(\ln f)'&=&\theta \nonumber\\
(\ln g)'&=&\theta\,(\ln u_*)'+(\ln u_*u_*')'\,,
\end{eqnarray}
where $\theta$ denotes the eigenvalue. Integration leads to the exact solution 
for the linear perturbation of the fixed point superpotential
\begin{equation}\label{delta1}
\delta u=C\, e^{\theta t}\,u_*^{\theta+1}\,u_*'\,.
\end{equation}
The allowed range of values for the exponents $\theta$ is determined using 
regularity conditions for the eigenperturbations.
To that end, we recall that the fixed point potential $u_*$ grows linearly 
with the field for large $\rho$, see \eq{largeR}, and hence
$\delta u\propto e^{\theta t} \rho^{\theta+1}$.
Furthermore, in the vicinity of the node we have \eq{small}, which for $c\neq 0$ 
leads to a finite $u'$ (meaning $0<u'<\infty$). We thus find
\begin{equation}
\delta u\propto e^{\theta t}\,(\rho-1)^{\theta+1}\,.
\end{equation}
Regularity of the perturbations requires non-negative integer 
values for the exponent $\theta+1$, reproducing \eq{theta1}. 

Note that this line of reasoning assumes analyticity of the perturbation
at the node which holds for all $c\neq 0$. For $c=0$, $u_*$ is non-analytical 
at \eq{zero} but $u_*^2$ instead is analytical and has a simple zero 
with finite $(u_*^2)'|_{u_*=0}$. Therefore we use \eq{delta1} to
relate the (regular) fluctuations of $u^2$ to $u_*^2$, leading to
\begin{equation}\label{delta2}
\delta u^2=C\, e^{\theta t}\,(u_*^2)^{\frac12(\theta+1)}\,(u_*^2)'\,.
\end{equation}
Again, analyticity implies that the exponent $(\theta+1)/2$ is a 
non-negative integer
\begin{equation}\label{theta2}
\theta=-1,1,3,5,7,\cdots
\end{equation}
Here we recognize the universal critical exponents of the $3d$ spherical 
model \cite{Litim:2001dt}. We stress, 
however, that this solution is not a proper fixed point solution in the usual sense 
because it is limited to field values with $\rho\ge 1$.

Finally we extend the analysis of  linear perturbations to those of the 
function $u^2$ and the scalar potential $v=\rho\,u^2$. We begin with 
$u^2=u_*^2+\delta u^2$. An analytical solution is found by using the identity 
$\delta u^2=2\,u_*\,\delta u$ together with \eq{delta1}, leading to
\begin{equation}\label{delta3}
\delta u^2=2C\, e^{\theta t}\,u_*^{\theta+2}\,u_*'\,.
\end{equation} 
Note that the degree in $u_*$ has increased by one unit. Employing the same 
reasoning as above  for $c\neq 0$, we conclude that the set of available eigenvalues is
 \begin{equation}\label{theta3}
\theta=-2,-1,0,1,2,3,\cdots
\end{equation}
Physically, the appearance of the  eigenmode with eigenvalue $-2$ is due to the mass
term squared, a term which on dimensional grounds is available in $u^2$ but not in $u$.  

Finally, using (\ref{dtv}), (\ref{delta3}) and \eq{eq:LPA15},  the linear eigenperturbations 
about the scalar potential  $v(t,\rho)=v_*+\delta v(t,\rho)$ are found as
\begin{equation}\label{delta4}
\delta v=2C\, e^{\theta t}\,u_*^{\theta+2}\left\{u_*+\,u_*'[1-u_*^2f(u_*^2)]\right\}\,.
\end{equation} 
Close to $u_*=0$, the term in square brackets reduces to 1, and the curly bracket 
becomes $u'_*$ which is finite at $u_*=0$. Therefore regularity of eigenperturbations 
again implies \eq{theta3}.  
In a non-supersymmetric scalar theory the potential is not constrained to be of the 
product form \eq{v} and 
an additional eigenvalue $-3$ becomes available related to redundant shifts of the potential. 

We conclude that supersymmetry is responsible for the absence of the redundant 
eigenvalue $-3$ in the scalar potential, and for relating  its two relevant 
eigendirections with eigenvalues $-1$ and $-2$ with the sole relevant eigendirection  
with eigenvalue $-1$ of the derivative of the superpotential.

\subsection{Wilson-Fisher fixed point}\label{WFFP}

It is interesting to clarify how the supersymmetric model and its fixed points fall 
back onto those of the $3d$ non-supersymmetric scalar theory in the same 
approximation \cite{Tetradis:1995br,Litim:1995ex,Litim:2002cf}. To that end, we consider 
the $4d$ supersymmetric $O(N)$ at finite temperature. The temperature is implemented 
using the imaginary time formalism which on the level of the flow equation amounts 
to the replacements \cite{Litim:1998yn,Berges:2000ew,Litim:2001up}
\begin{equation}\label{T}
\int_{-\infty}^{\infty}\frac{dq_0}{2\pi}f(q_0)\to T\sum_{n=-\infty}^\infty f(q_0=2\pi c_n T)\,.
\end{equation}
Here $2\pi c_n T$ denotes the $n$th Matsubara frequency with $c_n=n$ for bosons 
and $c_n=n+\frac12$ for fermions. The temperature imposes periodic (anti-periodic) 
boundary conditions for bosons (fermions) and, consequently, softly breaks 
global supersymmetry. Within a derivative expansion the relevant momentum 
integrals are performed analytically using the four-dimensional version of 
\eq {eq:LPA5} together with \eq{T} and the optimized momentum cutoff \eq{eq:LPA7}. 

\begin{figure}[t]
\begin{center}
\includegraphics[width=\columnwidth]{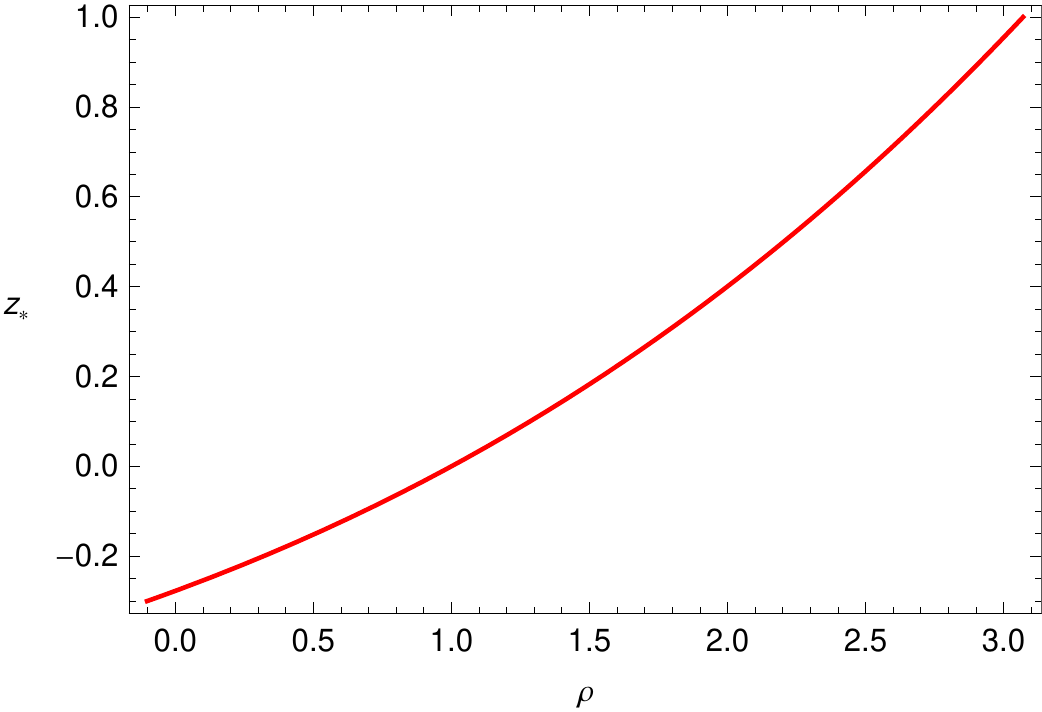}
\end{center}
\caption{\label{WF}The Wilson-Fisher fixed point solution $z_*(\rho)$ of \eq{dz}.}
\end{figure}

We are interested in the large-scale behavior $k/T\to0$.
Due to \eq{T}, all fermions and  bosons with a non-vanishing Matsubara mass will 
decouple from the system, except for the bosonic zero mode. In this limit, the 
$4d$ supersymmetric model undergoes a dimensional reduction to a $3d$ 
non-supersymmetric theory where all fermions have decoupled. In the large-$N$ limit, 
the RG equation for the potential of the remaining bosonic zero mode in LPA is given by
\begin{equation}
\label{dz}
\partial_t z=-2z+\rho\,z'-\frac{1-z}{(1+z)^2}\,z'
\end{equation}
where $z$ is related to the scalar field potential by $v(\rho)=\rho\,z(\rho)$. The 
key difference to the supersymmetric system studied previously is that the 
function $z$ is no longer constrained to be the square of a superpotential 
derivative $w'$. Relaxing this constraint allows for an additional fixed point 
solution, which follows from integrating \eq{dz} analytically. The general solution reads
\begin{equation}
\label{z}\frac{\rho-1}{\sqrt{z}}-\frac{\sqrt{z}}{1+z}-2\arctan\sqrt{z}=B(z\,e^{2t})
\end{equation}
where $B(z\,e^{2t})$ is fixed through initial conditions. The solution for 
negative $z$ is found by analytical continuation. In particular,  \eq{dz} has a 
Wilson-Fisher fixed point solution $z_*\neq 0$ with $z(\rho=1)=0$ corresponding 
to \eq{z} with $B=0$. The solution  extends over all $\rho$ with one unstable 
direction, see Fig.~\ref{WF}. The eigenperturbations $z=z_*+\delta z$ are found 
analytically leading to \eq{delta2} with the replacements $\delta u \to \delta z$ 
and $u^2_*\to z_*$. Hence, the universal eigenvalues are identical and given by \eq{theta2}. 

The similarities and differences between the Wilson-Fisher fixed point solution 
of the purely scalar theory and the $c=0$ `would-be' Wilson-Fisher fixed point 
of the supersymmetric partner theory can also be appreciated from the behavior 
at small and large fields. In fact, for $\rho\ge 1$, $z_*(\rho)$ is positive 
and related to the real superpotential by 
\begin{equation}
z_\ast(\rho)=w'_\ast(\rho)^2\,.
\end{equation}
In turn, $z_*(\rho)$ is negative for all $\rho<1$. Interestingly, this solution 
is still visible in the supersymmetric theory where it corresponds to a purely 
imaginary ``superpotential'' with 
\begin{equation}
w_\ast'(\rho)=\pm i \sqrt{-z_*(\rho)}\,.
\end{equation}
Hence, provided that a purely imaginary superpotential is meaningful in the 
supersymmetric theory, the $c=0$ solution can be extended to a valid supersymmetric 
Wilson-Fisher fixed point for all $\rho$. 
However, the structure of the Lagrangian imposed by supersymmetry implies
that the field-dependent fermion mass term is proportional to $w_\ast'$ and 
the Yukawa-type fermion-boson interaction proportional to $w_\ast''$ 
all become purely imaginary. Most importantly, a purely imaginary $w_\ast'$ 
for small fields implies that the scalar potential obeys 
$v_*(\rho)=\rho w_*'^2 < 0$ for all fields within $0<\rho<1$. Unbroken global supersymmetry 
requires that the dimensionful $V_k(\bar\varrho)$  remains positive for all fields 
and scales. In the infrared limit $k\to 0$, re-inserting powers of $k$, the dimensionful 
potential approaches $V(\bar\varrho)=64\pi^2\bar\varrho^3/N^2\ge0$. 
Hence, our results state that this potential can be approached arbitrarily close 
from within a phase with $O(N)$ symmetry and global supersymmetry.

\section{Conclusions}\label{Conclusion}

We have studied fixed points of supersymmetric $O(N)$ symmetric Wess-Zumino 
models in the limit of many components $N\to\infty$ in three dimensions with 
the help of the renormalization group. We have solved the theory analytically, 
showing that it displays a line of non-trivial fixed points solely parametrized 
by the exactly marginal linear superfield coupling. The fixed points are 
non-Gaussian, yet they display Gaussian exponents similar to the line of 
fixed points observed in the bosonic $(\phi^2)^3$ theory. The line 
of fixed points contains the Gaussian fixed point and therefore all 
fixed-points are continuously linked to the Gaussian one. With increasing 
superfield coupling, the line of fixed points bifurcates into two fixed point 
solutions, both of which terminate at a critical coupling \eq{cL2} below which 
no fixed point solutions exist which extends over all physical fields. One of these 
solutions has its minimum 
at $\rho_0=0$ the other at $\rho_0\neq0$. Interestingly, remnants of the 
non-Gaussian scaling exponents of the $3d$ spherical model \eq{theta2} become 
visible for asymptotically large superfield coupling. However, the fixed point 
solution does not extend over all fields in the supersymmetric 
case, except if the superfield potential becomes purely imaginary for small fields.

From a structural point of view, the main impact of global supersymmetry 
on the critical behavior in comparison with the purely scalar theory is 
summarized as follows. Firstly, for unbroken global supersymmetry the 
scalar potential has its minimum at vanishing field. Hence the irrelevant 
eigenmode with eigenvalue $-3$ corresponding to overall shifts in the 
potential is absent from the supersymmetric eigenvalue spectrum. Secondly, the quartic 
and sextic coupling of the scalar potential are no longer independent. 
Hence, in the supersymmetric theory criticality is achieved by tuning 
only one parameter as opposed to the tuning of two parameters in the 
corresponding purely bosonic theory. This is reflected in the sole 
negative eigenvalue for $u$ as opposed to the two negative eigenvalues 
for both $u^2$ and $v$. Finally, at the coupling $|c|=c_L$ \eq{cL2} the 
supersymmetric model shares similarities with the Bardeen-Moshe-Bander 
phenomenon in the bosonic theory \cite{David:1984we}. The logarithmic singularity 
observed in \cite{David:1984we} is superseded by a square-root behavior in the 
supersymmetric case, a difference which can be traced back to the underlying regularizations. 

The fixed point solutions discussed in this paper describe 
the phase transition for the breaking of the $O(N)$ symmetry.
Analyzing the pattern of symmetry breaking and the phase transition 
between symmetric and broken phases in more detail, 
and relating our findings with earlier studies based on gap 
equations is deferred to an upcoming publication. Furthermore, stepping back to finite $N$ 
we expect modifications to the above picture, both within the local potential approximation 
studied here and to higher order in the derivative expansion. For example, it is known that the $N=1$ model 
displays a superscaling relation linking the unstable direction with the anomalous 
dimension \cite{Gies:2009az,Synatschke:2010ub},  a behaviour which is quite different from 
the Ising universality class \cite{Litim:2010tt}. It will thus be interesting to see how 
these patterns generalize for supersymmetric $O(N)$ models with generic $N$.

\acknowledgments{Helpful discussions with Jens~Braun, Moshe~Moshe and 
in particular  Holger~Gies  are
gratefully acknowledged.  This work has been supported
by the Science and Technology Facilities Council (STFC)  
[grant number ST/G000573/1], the Studienstiftung des deutschen Volkes and 
the German Science Foundation (DFG) under GRK 1523 and grant Wi 777/10-1.}\\

\appendix

\section{Conventions} 
\label{sec:Conventions} 

Relevant
 symmetry relations and Fierz identities for Majorana spinors are  
 $\bar{\Psi}\chi = \bar{\chi}\Psi$, $\bar{\Psi}\gamma^{\mu} \chi = 
- \bar{\chi}\gamma^{\mu}\Psi$ and $\theta_k \bar{\theta}_l = 
- \frac{1}{2} (\bar{\theta}\theta)\mathds{1}_{kl}$.
One of the main features of the action is its invariance under
supersymmetry transformations. The latter are characterized by  the
supersymmetry variations $\delta_{\epsilon}\Phi^{i}$, generated by the
$\mathcal{N}=1$ fermionic generator  $\mathcal{Q}$.
We have
\begin{equation}
	\delta_{\epsilon}\Phi^{i}(x) = i \bar{\epsilon}_k\mathcal{Q}_k\Phi^{i}(x) \; 
	\mbox{with} \; 
	\mathcal{Q}_k = -i\partial_{\bar{\theta}_k} - \gamma^{\mu}_{kl}\theta_l\partial_{\mu}.
	\label{eq:variation}
\end{equation}
Thus,  (\ref{eq:variation}) leads to the
supersymmetry variations
\begin{equation}
	\delta \phi^{i} = \bar{\epsilon}\psi^{i},\; 
	\delta\psi^{i} = (F^{i} + i\slashed{\partial}\phi^{i})\epsilon\;
	\mbox{and}\;
	\delta F^{i} = i\bar{\epsilon}\slashed{\partial}\psi^{i}
	\label{eq:variationcomp}
\end{equation}
of the component fields.
The anticommuting sector of the superalgebra is given by the anticommutator of two supercharges
\begin{equation}
 \{\mathcal{Q}_k, \bar{\mathcal{Q}}_l\} = 2 \gamma^{\mu}_{kl} \partial_{\mu}.
 \label{eq:anticommutator}
 \end{equation}

 \begin{widetext}
\section{Superspace}
\label{sec:Derivation}

Following \cite{Synatschke:2008pv} we consider the action of the three-dimensional  supersymmetric $O(N)$
 model in the local potential approximation
\begin{equation}
\Gamma_k[\Phi^i] 
= \int\! d^3x\,\frac{d\theta_1 \, d\theta_2}{2i} \left(-\frac{1}{2} \Phi^i K
\Phi_i + 2 W_k(R)\right),\label{anctiontrunc}
\end{equation}
where $R = \frac{1}{2}\Phi^{i}\Phi_i$, $K = \frac{1}{2}(
\bar{\mathcal{D}}\mathcal{D} - \mathcal{D}\bar{\mathcal{D}})$ and $i = 1,...,N.$
We derive the flow equation  in the superspace $\mathbb{R}^{3|2}$ with  coordinates $z=(x, \theta_1, \theta_2)$. 
Furthermore, we introduce the abbreviation $\int dz \equiv \int
d^3x\,d\theta_1\,d\theta_2/(2i)$. In {Minkowski spacetime}
\cite{Synatschke:2010ub}, the Wetterich equation in superspace may be written in the form
\begin{align}
\partial_t \Gamma_k &=  \frac{i}{2} \!\int\!\! dz\, dz^{\prime}
(\partial_tR_k)_{mn} (z, z^{\prime})(G_k)_{nm} (z^{\prime}, z),\quad
t=\ln(k^2/\Lambda^2),\label{eq:wetterich}
\end{align}
 where $(R_k)_{mn}$ represents a supersymmetric regulator term and $(G_k)_{nm}$
 the connected Green's function.
According to
\cite{Synatschke:2008pv, Synatschke:2010ub}, we  now choose a
general regulator term {quadratic} in the superfields $\Phi^{i}$  and
{diagonal} with respect to the field indices:
\begin{equation}
\begin{split}
\Delta S_k &= \frac{1}{2} \int dz  \,\Phi^{i} R_{k, ij}(\mathcal{D}, \bar{\mathcal{D}})\Phi^j
= \frac{1}{2} \int dz  \,\Phi^{i}\left(2r_1 (-\partial_x^2, k)\delta_{ij}  
- r_2  (-\partial_x^2,k)K\delta_{ij}\right)\Phi^j. \label{eq:regulator2}	
\end{split}
\end{equation}
Notice that this regulator  conserves both the $O(N)$ symmetry and
supersymmetry.
The functional derivative with respect to a superfield is chosen according to the
 conventions $\frac{\vec{\delta}}{\delta\Phi^{j}(\tilde{z})} \int
 dz\,\Phi^{i}(z) = \delta^{i}_j$ with
 $\frac{\vec{\delta}\Phi^{i}(z)}{\delta\Phi^{j}(\tilde{z})} =
 2i\,\delta^{i}_j\,\delta(x-\tilde{x})\,\delta(\theta_2-\tilde{\theta}_2)\,
\delta(\theta_1-\tilde{\theta}_1)
 \equiv \delta^{i}_j\delta(z-\tilde{z})$.
Thus,
the second functional derivative  of the effective average action with respect
to the superfields reads
\begin{align}
\Gamma^{(2)}_{k, nm}(z, z^{\prime})
\equiv \frac{\overrightarrow{\delta}}{\delta\Phi^n(z)}\Gamma_k
\frac{\overleftarrow{\delta}}{\delta\Phi^m(z^{\prime})} 
 =\left[\big(-K + 2W_k^{\prime}(R)\big)\delta_{nm} + 2
 W_k^{\prime\prime}(R)\Phi_n\Phi_m\right]{(z)}\delta(z-z^{\prime}).
\label{eq:gammatwo}
\end{align}
Similarly, the second functional derivative $\Delta S_k^{(2)}(z, z^{\prime})$ of
the regulator term is given by 
 \begin{equation}
(R_{k})_{nm}(z, z^{\prime}) = \left[ 2r_1 - r_2K\right]{(z)}\,  
\delta_{nm}\,\delta(z-z^{\prime})
\label{eq:regulatorderivative}.
\end{equation}
Now we assume the superfields to be constant, i.e. $ \partial_x \Phi^{i}(x,
\theta)=0$, such that the regulator functions as well as the wave operator may
be simply  written in momentum space.
 However, note that the wave operator $K$ still contains derivatives with
 respect to the Grassmann coordinates and thus acts  on the adjacent
 delta functions.
Hence,  the flow of the effective average action may be written  
as
\begin{align}
\partial_t \Gamma_k &=\frac{i}{2}\int dz\, dz^{\prime}\,
(\partial_tR_k)_{mn}(z, z^{\prime})(\Gamma_k^{(2)} + R_k)_{nm}^{-1}(z^{\prime},z)\notag\\
 &= \frac{i}{2} \int\!d^3x\,\frac{ d\theta_1\, d\theta_2}{2i}\frac{d\theta_1^{\prime}\, 
d\theta_2^{\prime}}{2i}\!\!\int\!\! \frac{d^3p}{(2\pi)^3} (2\partial_t{r}_1
 - \partial_t{r}_2K){(p, \theta_1, \theta_2)} \,\delta_{mn}\,  2i\, \delta(\theta_2
- \theta_2^{\prime})\, \delta(\theta_1 -  \theta_{1}^{\prime} )\times\notag\\
&\quad\quad\big[(-hK_{(p, \theta_1^{\prime}, \theta_2^{\prime})} +
2\mathcal{W}^{\prime})\delta_{nm} + 2\mathcal{W}^{\prime\prime}\Phi_n\Phi_m\big]^{-1} 2i\,
 \delta(\theta_2^{\prime}- \theta_2)\, \delta(\theta_1^{\prime} -  \theta_{1}).
\label{eq:flowdetail}
\end{align}
We have thereby introduced the notation $\mathcal{W}^{\prime}(R) \equiv
W_k^{\prime}(R) + r_1$,  $h \equiv 1 + r_2$.
The  inverse  of the $N\times N$-matrix 
\begin{equation}
(M)^{-1}_{nm}\equiv (-hK + 2\mathcal{W}^{\prime})\delta_{nm} +
2\mathcal{W}^{\prime\prime}\Phi_n\Phi_m
\end{equation}
  is given by
 \begin{equation}
(M)_{nm} = \frac{(-hK + 2\mathcal{W}^{\prime})\delta_{nm} 
+ 2 \mathcal{W}^{\prime\prime}(\Phi^2\delta_{nm} - \Phi_n \Phi_m)}{4\left(h^2p^2  
+ \mathcal{W}^{\prime}(\mathcal{W}^{\prime} + 2\mathcal{W}^{\prime\prime}R)
- hK(\mathcal{W}^{\prime} + \mathcal{W}^{\prime\prime}R)\right)}, 
\label{eq:generalgreens}
\end{equation}
where we have  used the relation $K^2(p) = 4p^2$  
resulting from the action of 
 $K(p) = -\partial_{\theta} \partial_{\bar{\theta}} - (\partial_{\theta}p\!\!\!/\theta)
- (\bar{\theta}p\!\!\!/\partial_{\bar{\theta}}) - p^2 (\bar{\theta}\theta)$ 
on an arbitrary superfield. 
In order to eliminate the wave operator $K$ in the denominator of
(\ref{eq:generalgreens}), we multiply both the numerator and the denominator
with $\left[(h^2p^2 + \mathcal{W}^{\prime}(\mathcal{W}^{\prime} + 2\mathcal{W}^{\prime\prime}R))
 + hK(\mathcal{W}^{\prime} + \mathcal{W}^{\prime\prime}R)\right]$ and use again $K^2(p) = 4p^2$.
Thus we get
\begin{align}
(M)_{nm} 
&=-2\frac{h^2p^2(\delta_{nm}\mathcal{W}^{\prime} + \mathcal{W}^{\prime\prime}\Phi_n\Phi_m) 
- \mathcal{W}^{\prime}(\mathcal{W}^{\prime} 
+ 2\mathcal{W}^{\prime\prime}R)(\mathcal{W}^{\prime}\delta_{nm} 
+ \mathcal{W}^{\prime\prime}(\Phi^2\delta_{nm}-\Phi_n\Phi_m))}{4(h^2p^2 
- \mathcal{W}^{\prime 2})(h^2p^2 - (\mathcal{W}^{\prime} 
+ 2\mathcal{W}^{\prime\prime}R)^2)}\notag\\
&\quad - hK\frac{\delta_{nm}(h^2p^2 - \mathcal{W}^{\prime 2}) 
- 2\mathcal{W}^{\prime\prime}(\mathcal{W}^{\prime} 
+ \mathcal{W}^{\prime\prime}R)(\Phi^2\delta_{nm}-\Phi_n\Phi_m))}{4(h^2p^2 
- \mathcal{W}^{\prime 2})(h^2p^2 - (\mathcal{W}^{\prime} 
+ 2\mathcal{W}^{\prime\prime}R)^2)}\notag
\\
&\equiv\frac{-2f_{nm}\; - \; hKg_{nm}}{\mathcal{R}} \label{eq:inverseresult}
\end{align} 
with $\mathcal{R}= 4(h^2p^2 - \mathcal{W}^{\prime 2})(h^2p^2 
- (\mathcal{W}^{\prime} + 2\mathcal{W}^{\prime\prime}R)^2).$
For
\begin{equation}
(G_k)_{nm (p, \theta_1^{\prime}-\theta_1, \theta_2^{\prime}-\theta_2)} 
= (M)_{nm}(p,\theta_1^{\prime}, \theta_2^{\prime})\, 2i\, \delta(\theta_2^{\prime}
- \theta_2)\delta(\theta_1^{\prime}-\theta_1)
\label{eq:resultgreens}
\end{equation}
to be the Green's function it has to fulfill the defining relation
\begin{equation}
  \begin{split}
 &\int dz\, (G_k)_{mn}(\tilde{z},z)(\Gamma^{(2)}_k + R_k)_{np}(z, z^{\prime}) =
 \delta(\tilde{z}-z^{\prime})\delta_{mp}.
\label{eq:toshow}
\end{split}
\end{equation}
This can be shown by directly inserting the explicit expressions on the left
hand side and working out the contributions to different orders in $K$.

The flow equation is calculated by inserting the regulator  \eqref{eq:regulatorderivative} 
as well as the propagator \eqref{eq:resultgreens} into eq.
(\ref{eq:flowdetail}). Note that the regulator  $(R_k)_{mn}\propto 
\delta_{mn}$ is diagonal with respect to the field indices. Hence,  we simply
have evaluate the trace over the Green's function $(G_k)_{mm}$.  
This yields
\begin{align}
\partial_t\Gamma_k &= \frac{i}{2} \int\!\!d\theta_1\, d\theta_2 \,  
d\theta_1^{\prime}\,  d\theta_2^{\prime} \int\! \!d^3x \int\!\!\frac{d^3p}{(2\pi)^3} 
(2\partial_t{r}_1 - \partial_t{r}_2K){(p, \theta_1, \theta_2)}\, \delta(\theta_2
 - \theta_2^{\prime}) \, \delta(\theta_1 - \theta_1^{\prime})\notag\\
&\times\left(-2\frac{h^2p^2(N\mathcal{W}^{\prime} + 2\mathcal{W}^{\prime\prime}R) 
- \mathcal{W}^{\prime}(\mathcal{W}^{\prime} + 2\mathcal{W}^{\prime\prime}R)(N\mathcal{W}^{\prime}
 + 2(N-1)\mathcal{W}^{\prime\prime}R)}{4(h^2p^2 - \mathcal{W}^{\prime 2})(h^2p^2 
- (\mathcal{W}^{\prime} + 2\mathcal{W}^{\prime\prime}R)^2)}\right.\notag\\
&\left.\quad - hK\frac{N(h^2p^2 - \mathcal{W}^{\prime 2}) 
- 4(N-1)\mathcal{W}^{\prime\prime}(\mathcal{W}^{\prime} 
+ \mathcal{W}^{\prime\prime}R)R}{4(h^2p^2 - \mathcal{W}^{\prime 2})(h^2p^2 
- (\mathcal{W}^{\prime} + 2\mathcal{W}^{\prime\prime}R)^2)}\right)\delta(\theta_2^{\prime} 
- \theta_2) \, \delta( \theta_1^{\prime}- \theta_1).
\label{eq:flowequation}
\end{align}
Now, only terms {linear} in $K$ contribute to the flow of $\Gamma_k$ after
having integrated out the Grassmann variables. Those contributing terms lead
to a multiplying factor of $2i$.
  Thus the flow equation \eqref{eq:flowequation} simplifies to
\begin{align}
\partial_t\Gamma_k &= -i \int\!\!d^3x\, d\theta_1 \, d\theta_2\,
\partial_t{W}_k(R) \notag\\ &= \frac{1}{2}\int\!\!d^3x\, d\theta_1 \, 
d\theta_2  \int\!\!\frac{d^3p}{(2\pi)^3}\left((N-1) \frac{(\partial_t{r}_1 h - 
\partial_t{r}_2\mathcal{W}^{\prime})}{h^2p^2 - \mathcal{W}^{\prime 2}} +
\frac{\partial_t{r}_1 h - \partial_t{r}_2(\mathcal{W}^{\prime} +
2\mathcal{W}^{\prime\prime}R)}{h^2p^2 - (\mathcal{W}^{\prime} 
+ 2\mathcal{W}^{\prime\prime}R)^2}\right).
\label{eq:flowresultone}
\end{align}
Performing a Wick rotation of the zeroth component of the momentum, i.e. $p^0
\rightarrow ip^0_E$, $p^2\rightarrow -p^2_E$, we obtain the {Euclidean
version} of the flow equation (\ref{eq:flowresultone}).
Thus, the resulting flow equation in superspace reads
\begin{equation}
 \int_{x, \theta_1, \theta_2}\, \partial_t{W}_k(R)
= \frac{1}{2}\int_{x, \theta_1, \theta_2} 
\int\!\!\frac{d^3p_E}{(2\pi)^3}\left((N-1)\frac{(\partial_t{r}_1 h -
\partial_t{r}_2\mathcal{W}^{\prime})}{h^2p_E^2 + \mathcal{W}^{\prime 2}} 
+ \frac{\partial_t{r}_1 h - \partial_t{r}_2(\mathcal{W}^{\prime} 
+ 2\mathcal{W}^{\prime\prime}R)}{h^2p_E^2 + (\mathcal{W}^{\prime}  
+ 2\mathcal{W}^{\prime\prime}R)^2}\right).
\label{eq:flowresulttwo}
\end{equation}
Notice, that the truncation (\ref{eq:LPA3}) involved a superpotential of the 
form $2NW_k(R/N)$ instead of $2W_k(R)$. The corresponding flow equation may be easily
derived from the above result by performing the substitution
$W_k(R)\rightarrow NW_k(R/N)$ in (\ref{eq:flowresulttwo}). This yields the final
result
\begin{equation}
 \int_{x, \theta_1, \theta_2}\, \partial_t{W}_k(R/N)
= \frac{1}{2}\int_{x, \theta_1, \theta_2}  \int\!\!\frac{d^3p_E}{(2\pi)^3}\left(
\frac{(N-1)}{N}\frac{(\partial_t{r}_1 h -
\partial_t{r}_2\mathcal{W}^{\prime})}{h^2p_E^2 + \mathcal{W}^{\prime 2}} +
\frac{1}{N}\frac{\partial_t{r}_1 h - \partial_t{r}_2(\mathcal{W}^{\prime} +
2\mathcal{W}^{\prime\prime}R/N)}{h^2p_E^2 + (\mathcal{W}^{\prime} +
2\mathcal{W}^{\prime\prime}R/N)^2}\right).
\label{eq:flowresultthree}
\end{equation}   
 
\end{widetext}


\end{document}